\documentclass[traditabstract]{aa} 
\usepackage{graphicx}
\usepackage{txfonts}
\usepackage{natbib} 
\usepackage{xcolor} 
\bibpunct{(}{)}{;}{a}{}{,} 
\newcommand{\vc}[1]{\mbox{\boldmath{$#1$}}} 
\DeclareMathSymbol{\varGamma}{\mathord}{letters}{"00} 
\DeclareMathSymbol{\varOmega}{\mathord}{letters}{"0A} 

\begin{document}

 \title{Rapid growth of gas-giant cores by pebble accretion}
 \author{Michiel Lambrechts \inst{1} \and Anders Johansen\inst{1}}
 \institute{$^1$Lund Observatory, Department of Astronomy and Theoretical
 Physics, Lund University, Box 22100, Lund, Sweden\\
 \email{michiel@astro.lu.se}}
 \date{Received --- ; accepted ---}

\abstract{
The observed lifetimes of gaseous protoplanetary discs place strong constraints
on gas and ice giant formation in the core accretion scenario. The approximately
10-Earth-mass solid core responsible for the attraction of the gaseous envelope
has to form before gas dissipation in the protoplanetary disc is completed
within 1-10 million years. Building up the core by collisions between km-sized
planetesimals fails to meet this time-scale constraint, especially at wide
stellar separations. Nonetheless, gas-giant planets are detected by direct
imaging at wide orbital distances.  In this paper, we numerically study the
growth of cores by the accretion of cm-sized pebbles loosely coupled to the gas.
We measure the accretion rate onto seed masses ranging from a large planetesimal
to a fully grown 10-Earth-mass core and test different particle sizes. The
numerical results are in good agreement with our analytic expressions,
indicating the existence of two accretion regimes, one set by the azimuthal and
radial particle drift for the lower seed masses and the other, for higher
masses, by the velocity at the edge of the Hill sphere.  In the former, the
optimally accreted particle size increases with core mass, while in the latter
the optimal size is centimeters, independent of core mass.  We discuss the
implications for rapid core growth of gas-giant and ice-giant cores.  We
conclude that pebble accretion can resolve the long-standing core accretion
time-scale conflict.  This requires a near-unity dust-to-gas ratio in the
midplane, particle growth to mm and cm and the formation of massive
planetesimals or low radial pressure support.  The core growth time-scale is
shortened by a factor 30--1,000 at 5 AU and by a factor 100--10,000 at 50 AU,
compared to the gravitationally focused accretion of, respectively,
low-scale-height planetesimal fragments or standard km-sized planetesimals.
}


\maketitle
%

\section{Introduction}

The gas giants (Jupiter and Saturn) and ice giants (Uranus and  Neptune) in our
solar system consist of a dense rocky and/or icy core surrounded by a varying
degree of hydrogen and helium atmosphere \citep{Guillot_2005}.  The strong
positive correlation between stellar metallicity and exoplanet occurrence
\citep{Santos_2004,Fischer_2005} is also accompanied by a correlation between
stellar metallicity and the amount of heavy elements present in the exoplanetary
interior \citep{Guillot_2006,Miller_2011}, for objects in the gas giant mass
range between about 0.3 and 10 Jupiter masses ($M_{\rm J}$).  Additionally, a
careful statistical inspection of the planet candidates from the Kepler transit
survey reveals the evaporation and sublimation of the smaller ice and gas giant
planets to their naked cores as they get close to their host star
\citep{Youdin_2011}.  After their formation, the migration of these massive
planets in the later stages of the protoplanetary disc shapes the final
planetary system \citep{Walsh_2011}. However, reconstructing how ice and gas
giants form in the first place has proven to be challenging.

In the \emph{disc instability} scenario, gravitational instabilities in the
protoplanetary disc excite dense spiral arms which fragment directly into gas
giant planets \citep{Boss_1997}.  The \emph{core accretion} (or nucleated
instability) scenario requires the formation of a 10-Earth-mass ($M_\oplus$)
solid core, capable of holding on to a gaseous atmosphere.  When the envelope
reaches a mass comparable to the core mass, a run-away accretion of the
surrounding gas is triggered \citep{Mizuno_1980, Pollack_1996}.

Using the solar system as a template for the end result of planet formation is
challenged by direct imaging of planetary companions to A-stars at wide stellar
separations. The system HR 8799, for example, contains at least 4 planets
separated from their host star by $14.5$, $24$, $38$ and $68$ AU, confined by an
inner and outer debris disc \citep{Marois_2010}. Best estimates of the planetary
masses place them all in the gas-giant range. The presence of the debris discs
reveals that growth to planetesimals occurs at wide orbital distances as well.
Another example of a directly imaged gas-giant planet, $\beta$ Pictoris b
\citep{Lagrange_2010}, orbits the host star at approximately $10$ AU. Fomalhaut
b detected in reflected visble light \citep{Kalas_2008}, with an upper mass
below 1 M$_{\rm J}$ \citep{Janson_2012}, is located far from the central start
at approximately $120$ AU.  LkCa 15b is a newly discovered gas-giant planet of
about $6$ M$_{\rm J}$, likely caught in the epoch of formation, orbiting at
approximately $20$ AU around a young solar-like star, with an estimated age of
only $2$ Myr \citep{Kraus_2012}.

Formation of gas giants by direct gravitational collapse has been shown to be
problematic. At large distances from the host star, \cite{Kratter_2010} point
out that it becomes increasingly difficult to clump gas with masses below the
deuterium burning limit.  Additionally, at smaller stellar separations, gas
cools too slowly for the spiral arms to fragment into bound clumps
\citep{Matzner_2005, Rafikov_2005}.

On the other hand, gas-giant formation by core accretion suffers from
exceedingly long time-scales at wide stellar separations \citep{Dodson_2009,
Rafikov_2011}.  Observations of dust infra-red emission \citep{Haisch_2001,
Currie_2009} and disc accretion \citep{Jayawardhana_2006} limit the lifetime of
the  gaseous component of the protoplanetary disc to $10^{6\dots7}$ yr.
Classical core formation by runaway planetesimal accretion is believed to take
more than $10^7$ yr beyond 5 AU, where the planetesimal number densities are low
\citep{Goldreich_2004}.  Planetesimals ($>$ km) get gravitationally focused on
to the core, but this effect can be significantly reduced when scattering events
drive up the random velocity component of the planetesimals.  

The formation of planetesimals, larger\--than\--km\--sized solid bodies bound by
self-gravity, is problematic in its own way. While classically considered to be
the building blocks of both rocky planets and gas-giant cores, the formation of
solids this size remains difficult to explain both theoretically and
experimentally.  Particle growth beyond cm-sizes by coagulation is inefficient
\citep{Blum_2008, Brauer_2008, Windmark_2012} and radial drift time\--scales for
m\--sized boulders are as short as a hundred orbital time\--scales
\citep{Adachi_1976, Weid_1977}.  On the other hand, one can circumvent this
so\--called meter barrier with turbulence induced by the magnetorotational
instability \citep[MRI,][]{Balbus_1991}, which excites local pressure bumps,
ideal regions for dust particle trapping and growth
\citep{Whipple_1972,Johansen_2009_zonal}. In dead zones where the MRI does not
operate, streaming instabilities can destabilize the relative motion between gas
and particles \citep{Youdin_2005, Johansen_2007, Bai_2010} and lead to the
formation of dense filaments.  When the particle density becomes sufficiently
high, large Ceres-sized planetesimals are formed through gravitational collapse
\citep{Johansen_2007_nature}. The streaming instability benefits strongly from
increased disc metallicities \citep{Johansen_2009_met, Bai_2010}, explaining
partly the higher occurrence rate of exoplanets around higher metallicity stars.  

Instead of building up cores of ice and gas giants with planetesimals, we
investigate in this paper the accretion of smaller particles, coupled to the gas
on approximately orbital time-scales.  Dust continuum observations of young
circumstellar discs around low-mass pre-main-sequence stars show growth of the
dominant particle size to mm and cm sizes within less than 1 Myr
\citep{Testi_2003, Wilner_2005, Rodmann_2006}.  The dynamics of these small
particles is influenced by the presence of the surrounding protoplanetary gas
\citep{Weid_1977}, through Epstein drag \citep{Epstein_1924}.

While drag helps reducing the random velocities of large planetesimals ($\geq 1$
km), \citet{Rafikov_2004} carefully investigated analytically the effect of drag
on smaller bodies ($\leq 1$ km) as assumed products of a collisional cascade.
However, he did not consider particles coupled to the gas on
shorter-than-orbital time-scales, excluding the pebble-sized objects seen in T
Tauri discs. He finds that nearly all fragments settle to the midplane of the
nebula and that gas drag is efficient enough to prevent dynamical excitation,
making core formation possible within nearly $10^6$ yr, as was later confirmed
in coagulation models by \citet{Kenyon_2009}.  Accretion of smaller,
pebble-sized particles onto protoplanets was first investigated by
\citet{Johansen_Lacerda_2010}, who numerically found that pebbles are accreted
from the entire Hill sphere, the region roughly corresponding to the maximal
gravitational reach of the core.  They identify a prograde particle disc, which
could explain the spin periods of asteroids and preferential prograde spin of
large asteroids.  The influence of gas drag on the interaction of single small
bodies and low\--mass planets was explored by \cite{Ormel_2010}. Analytically,
they calculated that protoplanets starting from $\sim$$10^3$ km can efficiently
accrete  $\sim$cm-sized particles with impact parameters comparable to the
radius of the Hill sphere. \citet{Ormel_2011} further investigated the
protoplanet growth stage with a thorough toy model including fragments,
planetesimals and embryos and stressed the importance of the gas disc
properties, such as a reduced local headwind and turbulence for fast growth.
\citet{Perets_2011}  analytically investigated the coalescence of binary
planetesimals due to drag forces and commented on the possibility of growth
through this mechanism.  \citet{Lyra_2008} had already earlier ran full disc
models of pressure bumps formed near the edges of the deadzone. After merely
$200$ orbits, they observed bound embryos with masses similar to the planet
Mars, consisting of pebble-sized particles. 

In this paper, we investigate core growth from a seed mass by gas-drag-aided
capture of cm-sized pebbles.  In Section 2, we describe the physics included in
the shearing coordinate frame used to numerically model the growth of the core.
In Section 3 we present the results from our simulations and analyse the
accretion rates for various core masses. We compare our results to analytic
expressions capturing the essential physics underlying the phenomena at hand,
namely the sub-Keplerian gas velocity, the particle size, the Keplerian shear
and the gravitational pull from the seed core. The effect of local changes in
the pressure gradient are analysed and we present the effect of including the
backreaction of the particles on the gas flow. By extrapolating the measured
accretion rates, we discuss the formation of gas and ice giant cores and derive
a characteristic time-scale for core formation by pebble accretion in Section 4.
We discuss the approximations made in this paper and the limitations of our
model in Section 5. Finally, in Section 6, we conclude that pebble accretion can
explain rapid gas and ice giant formation in the core accretion scenario, even
at wide stellar separations.   

\section{Physical model}

The growth of a gas giant's core occurs in a protoplanetary disc, a gaseous disc
in the process of accreting onto the young star.  Based on the mass distribution
in the solar system and assuming a mean gas-to-dust ratio or metallicity of 
\begin{equation}
  Z= \frac{\Sigma_{\rm p}}{\Sigma} = 0.01,
\end{equation}
with $\Sigma_{\rm p}$ and $\Sigma$ denoting the solid (dust+ice) and gas column
densities, \citet{Hayashi_1981} constructed the minimum mass solar nebula
(MMSN).  He found the radial dependence of the gas column density to be
\begin{equation} 
  \Sigma = 1700 \left( \frac{r}{\rm{AU}} \right)^{-3/2} \rm{ g\, cm}^{-2},
\end{equation}
with the orbital radius $r$ within $0.35$-$36$ AU. The thin disc is
characterized by the ratio of the gas scale height $H$ to the orbital distance 
\begin{equation} 
  \frac{H}{r} = \frac{c_{\rm s}}{v_{\rm K}} \approx 0.033 \left( \frac{r}{ {\rm
  AU}} \right)^{1/4},
\end{equation} 
with $c_{s}$ the sound speed of the gas and $v_{\rm K}$ the Keplerian velocity, \begin{equation}
  v_{\rm K} = r \varOmega_{\rm K} = \left( \frac{GM}{r} \right)^{1/2}.
\end{equation} 
Here $\varOmega_{\rm K}$ is the Keplerian frequency.  

\begin{figure}
  \centering
  \includegraphics{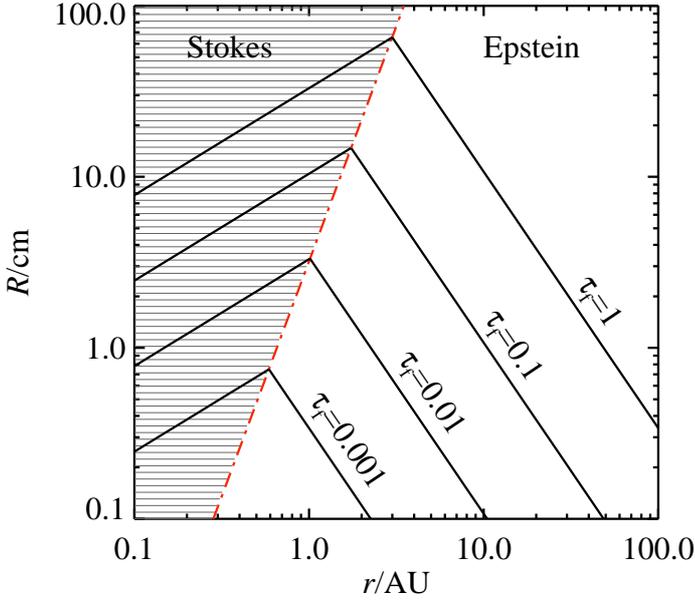}
  \caption{The physical response of a particle reacting to the gas flow is set
  by the friction time $t_{\rm f}$. A particle of given size $R$ has a
  dimensionless friction time $\tau_{\rm f} = \varOmega_{\rm K} t_{\rm f}$ that
  depends on the orbital distance $r$. The red dash-dotted line marks the
  distance at which the particle size equals 9/4 of the mean free path $\lambda$
  of molecular hydrogen in the MMSN.  Particles with $R>(9/4)\lambda$ are
  located in the dashed region, and experience Stokes drag as opposed to Epstein
  drag. For the curves in the Stokes regime, we have ignored the transition into
  the non-linear Stokes regime, applicable for large particles close to the host
  star. }
  \label{fig:particle_size}
\end{figure}

Solids with radii smaller than the local mean free path of the gas, $R \leq
(9/4) \lambda$, are in the Epstein regime of gas-particle coupling
\citep{Epstein_1924}.  They react on a friction time-scale $t_{\rm f}$ to
changes between the relative particle velocity $\vec{v}$ and the local gas
velocity $\vec{u}$,
\begin{eqnarray}
  \vec{\dot v}_{\rm drag} = - \frac{1}{t_{\rm f}} \left( \vec{v} - \vec{u}\right) =
  -\frac{\rho c_{\rm s}}{\rho_\bullet R} \left( \vec{v} -
  \vec{u}\right) ,   
\end{eqnarray}
where $R$ and $\rho_\bullet$ are the radius and material density of the
particle, while $\rho$ is the local gas density. For particles in the vicinity
of the midplane, with $z < H$, one can assume $\rho H \approx \Sigma  /
\sqrt{2\pi}$, so that the particle size $R$ in the MMSN can be recovered from
its dimensionless friction time 
\begin{eqnarray}
\tau_{\rm f} = \varOmega_{\rm K} t_{\rm f}
\end{eqnarray}
(also known as the Stokes number) as 
\begin{equation}
  R = 60 \rm{\, cm \, } \tau_{\rm f} \left( \frac{\rho_\bullet}{2
  \rm{\, g\, cm}^{-3}}\right)^{-1} \left( \frac{ {\it r}}{\rm{AU}} \right)^{-3/2}.
\end{equation}
Figure \ref{fig:particle_size} shows the relation between the orbital radius and
the particle radius for different dimensionless friction times. Around 10 AU, a
dimensionless friction time of $\tau_{\rm f} =0.1$ corresponds to cm-sized
particles, which we will refer to as pebbles.  Close to the star, the gas
density increases sufficiently for the particles to enter the Stokes drag
regime, where $\tau_{\rm f}^{\rm (S)} =  (4/9) (R/\lambda) \tau_{\rm f}$ scales
as $\propto$$r^{5/4}$. For a more complete description of different drag
regimes, see e.g.  \citet{Rafikov_2005} or \citet{Youdin_2008}. 

\begin{figure}
  \centering
  \includegraphics{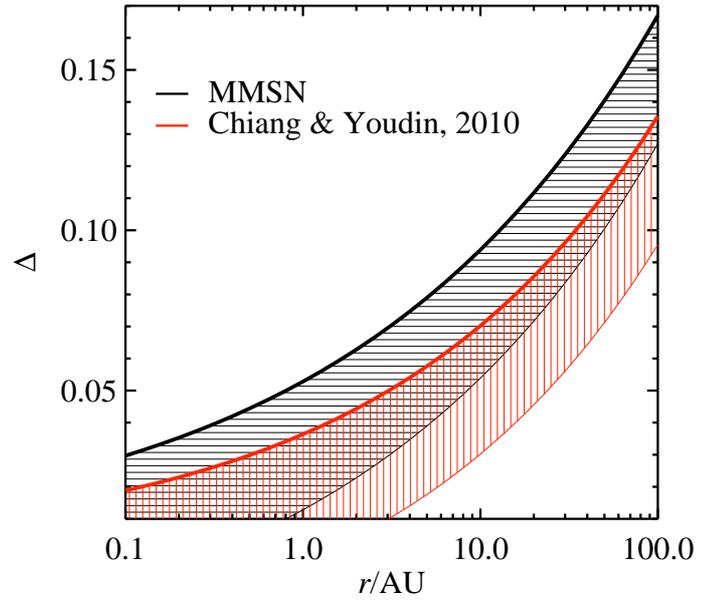}
  \caption{Deviation $\Delta$ of the orbital velocity of a gas element with
  respect to an object orbiting with the full Keplerian frequency, normalized by
  the local sound speed, is plotted as function of orbital radius $r$ in AU. The
  bold black line represents the traditional MMSN scaling, while the bold red
  line corresponds to the adapted MMSN as presented in \citet{Chiang_2010}.  The
  shaded area connecting to thin curves indicates the effect of a strong
  pressure bump of strength $\delta \Delta = -0.04$. The adopted standard value
  of $\Delta = 0.05$ in this paper, is accurate in a region around $5$ AU, even
  without a strong pressure bump.}
  \label{fig:headwind} 
\end{figure}

The gas component of the protoplanetary disc moves with a sub-Keplerian mean
velocity, since the force due to the the solar gravity is reduced by the
radially outwards pointing gas pressure force. The azimuthal velocity difference
$\Delta = \Delta u_\phi/c_{\rm s}$ between the mean gas flow and a pure
Keplerian orbit is given by
\begin{eqnarray}
  \Delta =\eta \frac{ v_{\rm K}}{c_{\rm s}} 
  = -\frac{1}{2}\frac{c_{\rm s}}{v_{\rm K}} \frac{\partial \ln(P)}{\partial
  \ln(r)}, 
  \label{eq:headwind}
\end{eqnarray}
where $P=\rho c_s^2$ is the gas pressure and $\eta$ is a measure of the gas
pressure support \citep{Nakagawa_1986}.  In the MMSN, $\Delta$ has a weak radial
dependency, $ \Delta \approx 0.05 \left( r/{\rm AU} \right)^{1/4} $, as can be
seen in Figure \ref{fig:headwind}.  However, comparison of the MMSN model with
observed protostellar accretion discs, \citep[e.g.][]{Bell_1997} and studies of
solar nebula metallicities \citep{Lodders_2003} have prompted updated MMSN
models, with a less steep pressure gradient, $ \Delta = 0.036 \left( r/{\rm AU}
\right)^{2/7}$  \citep{Chiang_2010}, as illustrated in Figure
\ref{fig:headwind}.

The turbulent nature of an accreting protoplanetary disc can result in local
pressure maxima \citep{Johansen_2009_zonal,Fromang_2009}.  As can be seen from
Eq.\ (\ref{eq:headwind}) these pressure bumps can locally reduce the headwind
the pebbles experience.  Reductions by $\delta \Delta \approx -0.02 $ are seen
in shearing box simulations of the MRI \citep{Johansen_2009_zonal,Fromang_2009}
and global simulations \citep{Lyra_2008_global_pressure}.  We have illustrated
the effect of a strong pressure bump, with $\delta \Delta \approx -0.04$, in
Figure \ref{fig:headwind}.

Since particles face a headwind, they will drift radially and azimuthally as
\begin{eqnarray}
  v_r    &=&  -2 \frac{\tau_{\rm f}}{\tau_{\rm f}^2+1} \eta v_{\rm K},\\
  v_\phi &=& - \frac{1}{\tau_{ \rm f}^2+1} \eta v_{\rm K},
\end{eqnarray}
as shown by \citet{Weid_1977} and \citet{Nakagawa_1986}.  The total relative
velocity between the particle and the core in pure Keplerian rotation is
\begin{equation}
  \Delta v = \frac{\sqrt{4\tau_{\rm f}^2 +1}}{\tau_{\rm f}^2+1} \eta
  v_{\rm K},
\end{equation}
which is well approximated by $\Delta v \approx \eta v_k$, since the particle
sizes we consider, $\tau_{\rm f} = ( 0.01,0.1,1 )$, give us $\Delta v / (\eta
v_{\rm K}) = ( 1.0,1.0,1.1 )$.

Particles settle in the vertical direction (perpendicular to the orbital plane).
The particle scale height $H_{\rm p}$ is a  balance between midplane-directed
gravity and turbulent diffusion parametrized by the coefficient $\delta_{\rm t}$
\citep{Youdin_2007},
\begin{equation}
  \frac{H_{\rm p}}{H} \approx \sqrt{\frac{\delta_{\rm t}}{\tau_{\rm f}}} \approx
  0.01,
\end{equation}
where we will make the approximation that this holds for the pebble size range
we consider ($\tau_{\rm f}=0.01$-$1$). The turbulence generated by the streaming
instability \citep{Youdin_2005} self-regulates the particle midplane density to
equal the gas density, independent of particle size. Since $Z (H_{\rm p}/H)^{-1}
\approx \rho_{\rm p}/ \rho \approx 1$, a near-unity midplane density in a
protoplanetary disc with metallicity $Z =0.01$ sets the particle scale height to
be $H_{\rm p}/H =0.01$.  For turbulence generated through the MRI, a value of
$\delta_{\rm t} = 0.001$ would give a ten times higher particle scale height,
$H_{\rm p}/H \approx 0.1$, for particles of friction time $\tau_{\rm f} = 0.1$.

\begin{figure}
  \centering
  \includegraphics{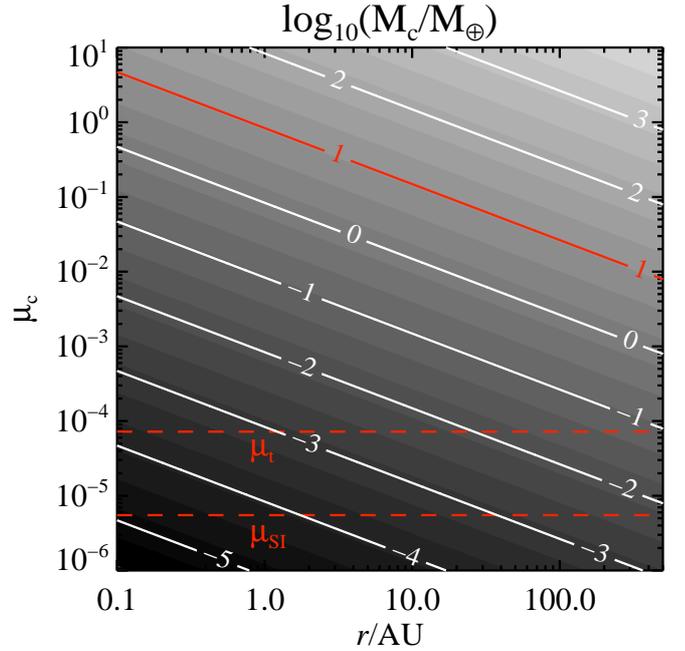}
  \caption{Contour lines mark the core mass $M_{\rm c}$ (in units of Earth
  masses M$_\oplus$), as function of the dimensionless mass unit $\mu_{\rm
  c}=(GM_{\rm c})/(\varOmega_{\rm K}^2 H^3)$ and orbital radius $r$ in AU.  The
  red contour line indicates the assumed minimal mass of a gas-giant core. The
  horizontal red dashed lines indicate the expected initial seed core mass from
  the streaming instability after planetesimal formation by gravitational
  collapse, $\mu_{\rm SI}$, and the transition mass $\mu_{\rm t}$ (see Section
  3.2).}
  \label{fig:muvsr}
\end{figure}

The aim of this paper is to investigate accretion onto cores of various masses,
ranging from the expected initial masses of planetesimals to estimated final
core masses of gas-giant planets.  The core mass, or more precisely its
gravitational parameter $GM_{\rm c}$, is non-dimensionalized as
\begin{equation}
  \mu_{\rm c} = \frac{GM_{\rm c}}{\varOmega_{\rm K}^2 H^3} 
  = \frac{M_{\rm c}}{M_\odot} \left( \frac{H}{r} \right)^{-3} \propto r^{-3/4},
\end{equation}
with $G$ the gravitational constant and $M_\odot$ the stellar mass.  Figure
\ref{fig:muvsr} relates the dimensionless core mass $\mu_{\rm c}$ to the orbital
radius in the MMSN.  Given the core mass, we can assign it an uncompressed
radius of
\begin{equation}
  R_{\rm c} = 890 \left( \frac{\rho}{2 \rm{ g/cm}^{3}} \right)^{-1/3}\left(
\frac{M_{\rm c}}{ 10^{-3} M_\oplus} \right)^{1/3} \rm{ km.}
\end{equation}
The critical core mass for runaway accretion of a gaseous envelope is
approximately $10$ M$_\oplus$ \citep{Mizuno_1980}, only weakly dependent on the
orbital radius outside the terrestrial planet region \citep{Rafikov_2006}. At
$5$ AU this mass corresponds to $\mu_{\rm c} \approx 1$, as can be seen in
Figure \ref{fig:muvsr}.  However, as \citet{Hori_2011} point out, if the
envelope can be significantly polluted by heavy elements from the accretion of
icy bodies, the critical core mass will be reduced by up to two orders of
magnitude.

Planetesimals are believed to have initial sizes in the $100-1000$ km region
\citep{Johansen_2007_nature,Morbi_2009}. \citet{Johansen_2012} find the
characteristic clump mass by streaming instabilities to be $\mu_{\rm SI} \approx
5 \times 10^{-6}$ (see Figure \ref{fig:muvsr}). 

The dynamical equations of the particles,
\begin{eqnarray}
  \frac{d \vc v}{dt} =
  -2 \vc \varOmega_{\rm K} \times \vc v + 3 \varOmega_{\rm K}^2 x \vc e_x 
  + \vc g_{\rm c} - \frac{1}{t_{\rm f}} \left( \vc v - \vc u \right),
  \label{eq:particles}
\end{eqnarray}
are solved with the Pencil Code\footnote{The Pencil Code can be freely obtained
at \\ {\tt http://code.google.com/p/pencil-code/}.} in the shearing box
approximation \citep{Goldreich_1980, Brandenburg_1995}.  A Cartesian coordinate
system is placed rotating at an arbitrary, but fixed orbital distance, with
Keplerian frequency $\varOmega_{\rm K}$. The $x$-axis points radially outwards,
the azimuthal direction corresponds to the $y$-coordinate and the vertical
$z$-direction is perpendicular to the midplane.  The motion of the particles is
described by Eq.\,(\ref{eq:particles}), which includes the acceleration due to
the core placed in the centre of the frame and the self-gravity of the particles
solved for through the Poisson equation, 
\begin{equation}
\nabla \cdot \vc  g_{\rm c} = 4\pi G \rho_{\rm p},
\end{equation}
Additionally, it includes the drag force term $-\frac{1}{t_f} \left( \vc v - \vc
u \right)$, a term balancing the linearized gravity and the centrifugal force
$3\varOmega_{\rm K}^2 x \vc e_{x}$ and the Coriolis force $-2 \vc \varOmega_{\rm
K} \times \vc v$.  

We perform simulations both with and without the backreaction term of the
particles on the gas, with gas backreaction time $(\rho_{\rm p} / \rho)^{-1}
t_{\rm f}$. The momentum equation for the fluid elements,
\begin{eqnarray}
  \frac{\partial \vc u}{\partial t} + \vc u \cdot \nabla \vc u
  &=& -2 \vc\varOmega_{\rm K} \times \vc u + 3\varOmega_{\rm K}^2 x \vc
  e_{x} -\varOmega_{\rm K}^2z \vc e_z \nonumber \\
  & & -\frac{\nabla P}{\rho_{\rm g}}
  + \frac{\rho_{\rm p}}{\rho_{\rm g} t_{\rm f} } \left( \vc v - \vc u \right),
  \label{eq:gas}
\end{eqnarray}
includes the pressure gradient term, $ - (1/\rho_{\rm p}) \nabla P$, and
vertical gravity $-\varOmega_{\rm K}^2z$.  The continuity equation for the
gaseous component of the protoplanetary disc is given by
\begin{eqnarray}
  \frac{\partial \rho}{\partial t} + \nabla \cdot (\rho \vc u) = &0&,
  \label{eq:gas_cont}
\end{eqnarray}
and we use artificial hyperdiffusivity for the gas to dissipate energy on the
smallest scales.

We solve these equations non-dimensionalized by the Keplerian frequency
$\varOmega_{\rm K}$, the scale height $H$ of the gas disc and $\rho_0$, the gas
midplane density. This has the benefit that, when interpreting the normalized
results, the orbital dependency is nearly fully recovered from these parameter's
MMSN orbital scalings (e.g. Figure \ref{fig:particle_size}, \ref{fig:headwind},
\ref{fig:muvsr}).  The numerical results are  however not completely scale-free.
When solving for self-gravity with the Poisson equation, we must set 
\begin{equation}
  \varGamma = \frac{4\pi G \rho_0 }{ \varOmega_{\rm K}^2}, 
\end{equation}
the non-dimensionalised form of the gravity constant $G$, as an initial
condition. It shows only a weak dependency on the orbital radius, 
\begin{equation}
  \varGamma \approx 0.04 \left(\frac{r}{3\rm{ AU}}\right)^{1/4},
\end{equation}
in the MMSN.  We therefore fix $\varGamma = 0.04$ for the remainder of this
paper.

\begin{table*}
  \caption{Characterizing parameters of all simulations used in this paper. All
  simulations are performed with $128^3$ grid cells resolution in stratified
  shearing boxes, with particle scale height $H_{\rm p}/H =0.01$ and
  metallicity $Z=0.01$. The first column gives the name of the simulation,
  followed by the characterizing parameters: the core mass $\mu_{\rm c}$,
  particle size $\tau_{\rm f}$, headwind parameter $\Delta$ and side length $L$
  of the cubic simulation domain. The last column indicates whether the simulation
  includes the backreaction (BR) term of the particles on the gas, or not.}
  \label{table:sims} 
  \centering  
  \begin{tabular}{c c c c c c} 
    \hline\hline  
    name & $\mu_{\rm c}$ & $\tau_{\rm f}$  &  $\Delta$ & $L$/$H$ & BR \\ 
    \hline 
    {\tt 1e-6\_0.01 / 1e-6\_0.1 }& $10^{-6}$ & $0.01/0.1$  &  0.05  & 0.01 & No\\
    {\tt  1e-6\_0.1\_0.03} & $10^{-6}$ & $0.1$  &  0.03  & 0.01 & No \\
    {\tt 2.5e-6\_0.1\_0.03 / 2.5e-6\_0.1 / 2.5e-6\_0.1\_0.07}& $2.5 \times 10^{-6}$ & $0.1$
    &  0.03 / 0.05 / 0.07  & 0.02 & No\\
    {\tt 1e-5\_0.01 / 1e-5\_0.1 / 1e-5\_1.0}& $10^{-5}$ & $0.01/0.1/1$  &  0.05
    & 0.04 & No\\
    {\tt 1e-5\_0.1\_0.03  / 1e-5\_0.1\_0.07}& $10^{-5}$ & $0.01/0.1/1$  &
    0.03/0.07  & 0.04/0.01 & No\\
    {\tt 1e-4\_0.01 / 1e-4\_0.1 / 1e-4\_1.0} & $10^{-4}$ & $0.01/0.1/1$  &  0.05
    & 0.128 & No\\
    {\tt 1e-3\_0.01 / 1e-3\_0.1 / 1e-3\_1.0} &  $10^{-3}$ & $0.01/0.1/1$  &
    0.05  & 0.32 & No\\
    {\tt 1e-3\_0.1\_b} & $10^{-3}$ &   $ 0.1 $  &  0.05  & 0.2 & Yes\\
    {\tt 1e-2\_0.01 / 1e-2\_0.1 / 1e-2\_1.0}& $10^{-2}$ & $0.01/0.1/1$  &  0.05
    & 0.64 & No \\
    {\tt 1e-1\_0.01 / 1e-1\_0.1 / 1e-1\_1.0}& $10^{-1}$ & $0.01/0.1/1$  &  0.05
    & 1.28 & No\\
    \hline
  \end{tabular}
\end{table*}

All simulations are performed in a three-dimensional shearing box, with a fixed
particle scale height of $H_{\rm p} / H = 0.01$.  Run parameters for all
simulations used in this paper can be inspected in Table \ref{table:sims}.  The
core is fixed in the origin of the coordinate system, a valid approximation for
the range of gas-decoupled seed masses we cover.  When the escape velocity
$v_{\rm esc}$ from the surface of the core is small compared the sound speed,
variations in the gas density can be ignored, as can be seen from the
hydrostatic equilibrium of an isothermal gas,
\begin{equation}
  \frac{v_{\rm esc}^2}{c_{\rm s}^2} \approx \frac{\partial \ln \rho}{\partial \ln r}.
\end{equation}
We argue this approximation holds up to the largest cores we consider in Section
5.  All simulations, with the exception of {\tt 1e-3\_0.1\_b} (see Table
\ref{table:sims}), do include the gas drag on the particles, but lack the
backreaction from the particles of the core on the gas. When omitting the
backreaction term, the gas velocity equals to the sub-Keplerian velocity, $u_y =
-\eta v_{\rm K}$.  However, simulation {\tt 1e-3\_0.1\_b} includes the
particle's backreaction on the gas and follows the numerical scheme discussed in
\citet{Youdin_Johansen_2007}. When including the backreaction term, we also turn
on the vertical gravity force for the particles, $-\varOmega_{\rm K}^2z \vc e_z
$.

All runs have sheared periodic boundary conditions in the radial direction, but
particles crossing azimuthal boundaries get removed from the simulation domain
(with exception of {\tt 1e-3\_0.1\_b}), in order to avoid accretion of particles
already focused from their first passage past the core.  We have also run
simulations including collisions, with the scheme discussed in
\citet{Johansen_2012}, and found no measurable difference in the accretion rates
on the seed core masses. We therefore omit collisions from the simulations in
this paper. Implications and limitations of the simulation set up are further
discussed in \mbox{Section 5}.

\section{Results}

Inspection of the particle's momentum equation, Eq.\,(\ref{eq:particles}),
reveals an important length scale. The Hill radius, 
\begin{equation} 
  r_{\rm H} = \left( \frac{GM_{\rm c}}{3\varOmega^2} \right)^{1/3},
\end{equation}
is set by the gravitational competition between the acceleration towards the
core and the stellar tidal field in the radial direction.  At a separation
$r_{\rm H}$ from the core, the orbital time around the core approximately equals
the orbital time around the star, $2\pi \varOmega_{\rm K}^{-1}$.  The Hill
sphere's radius grows linearly with the orbital radius $r_{\rm H} \propto r$,
placing more material in the gravitational region of influence of the core.

\subsection{Drift accretion}
When ignoring the stellar tidal field and the Coriolis force, the Bondi
radius\footnote{Note that we define the Bondi radius with the square of the
relative velocity between core and particle in the denominator, and not the
sound speed squared, which is also found in the literature.}
\begin{equation}
  r_{\rm B} =  \frac{GM_{\rm c}}{\Delta v^2},
\end{equation}
marks the outer point at which particles approaching the core with relative
velocity $\Delta v$ get significantly gravitationally deflected \citep[$\gtrsim
1$ rad, e.g.][]{Binney_1987}.  Ignoring the stellar tidal field is a valid
approximation before the core mass grows to the point where the Bondi radius
becomes comparable to the Hill radius (see Section 3.2), and we can associate a
core mass, 
\begin{equation}
  M_{\rm t} = \sqrt{\frac{1}{3}} \frac{\Delta v^3}{G\varOmega_{\rm K}},
\end{equation}
with this transition.  Masses with $M_{\rm c} < M_{\rm t}$ are in the
\emph{drift} regime and pebbles embedded in the gaseous disc approach the core
with a mean velocity comparable to the gaseous headwind the core experiences,
$\Delta v \approx \eta v_{\rm K}$.

Depending on the balance between the gravitational attraction of the core and
the drag force the pebbles experience, a particle can be pulled from the mean
gas flow and accreted if it dissipates enough energy while being deflected.
When the gas-free core-crossing time associated with the Bondi radius,
\begin{equation}
  t_{\rm B} = \frac{r_{\rm B}}{\Delta v},
  \label{eq:Bondi_time}
\end{equation}
is similar to the friction time $t_{\rm f}$, the drag force will cause all
pebbles within the Bondi radius to spiral inwards. However the effective
accretion radius, the drift radius $r_{\rm d}$, shrinks with respect to the
Bondi radius, when $t_{\rm B} \approx t_{\rm f}$ is not satisfied.

\begin{figure}
  \centering
   \includegraphics{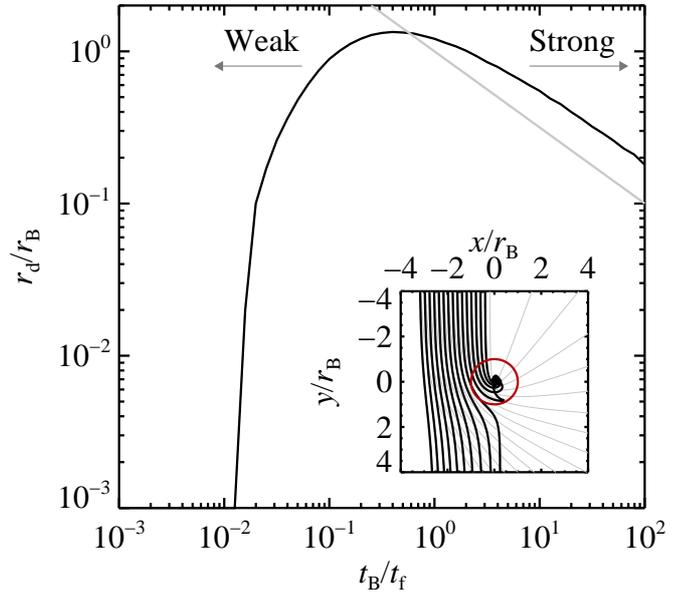}
   \caption{ Accretion efficiency in the weak and strong coupling regime.  When
   the Bondi time $t_{\rm B} = GM/\Delta v^3$ is equal to the particle's
   friction time $t_{\rm f}$, the drift accretion radius $r_{\rm d}$ peaks and
   equals the Bondi radius $r_{\rm B}$. For a particle of fixed size, the ratio
   $t_{\rm B}/t_{\rm f}$ on the horizontal axis increases as the core mass grows
   in time.  When particles are strongly coupled to the gas ($t_{\rm B} > t_{\rm
   f}$), with respect to the gravitational attraction of the core, the drift
   radius decreases as $r_{\rm d} \sim \left(t_{\rm B}/t_{\rm s} \right)^{-1/2}$
   (the analytical scaling of Eq.\,(\ref{eq:strong_couple_limt}) is indicated
   with a full grey line). Near $t_{\rm B}/t_{\rm f} \approx 10^{-2}$ the drift
   radius rapidly decreases. The inset shows particle trajectories (grey curves)
   in this regime, which can be compared with those at $t_{\rm B}/t_{\rm f} =1$
   (black curves). Where the former are simply gravitationally deflected, in the
   latter case we see that particles inside the Bondi radius (marked by a red
   circle) are accreted by the central point source.}
   \label{fig:drift_radius}
\end{figure}

When $t_{\rm B} >  t_{\rm f}$, the particle under consideration is
\emph{strongly coupled} to the gas. In this limit, only grazing particles
deflected on time-scales shorter than the friction time get pulled out of the
flow. If we let $g$ denote the gravitational attraction due to the core's mass,
the condition
\begin{equation}
  t_{\rm g}  = \frac{\Delta v}{g} < t_{\rm f}
  \label{eq:strongcouple}
\end{equation}
needs to be satisfied for accretion to occur. Since the deflection time $t_{\rm
g}$ is given by $ \Delta v r^2/ (GM_{\rm c}) = \left( r/r_{\rm B} \right)^2
t_{\rm B}$, the effective drift accretion radius is given by
\begin{equation}
  r_{\rm d} = \left(\frac{t_{\rm B}}{t_{\rm f}} \right)^{-1/2} r_{\rm B},
  \label{eq:strong_couple_limt}
\end{equation}
in the strong coupling limit. This radius corresponds to the settling radius in
\citet{Ormel_2010} and is also equal to the radius found by \citet{Perets_2011}
where drag forces shear apart bound binaries in the Epstein regime.  We verified
this power law by numerically integrating orbits of test particles in the 2-body
problem including drag,
\begin{eqnarray}
  \frac{\partial v_x/ \Delta v}{\partial t/t_{\rm B}} &=& 
  - \left( \frac{r_{\rm B}}{r} \right)^3 \frac{x}{r_{\rm B}} 
  - \frac{t_{\rm B}}{t_{\rm f}} \frac{v_x}{\Delta v},
  \\
  \frac{\partial v_y/ \Delta v}{ \partial t/t_{\rm B}}  
  &=& - \left( \frac{r_{\rm B}}{r} \right)^3 \frac{y}{r_{\rm B}} 
  - \frac{t_{\rm B}}{t_{\rm f}} \left( \frac{v_y}{ \Delta v} -1 \right),   
\end{eqnarray}
where we non-dimensionalized the particle equation of motion, ignoring disc
dynamics. This is a valid approximation in the \emph{drift} regime, where
$t_{\rm B} \ll \varOmega_{\rm K}^{-1}$.  Sample orbits can be investigated in
the inset of Figure \ref{fig:drift_radius}, which shows the maximal
particle-core separation leading to capture. The drift radius for strongly
coupled particles falls of as $\propto \left( t_{\rm B}/t_{\rm f}
\right)^{-1/2}$ as predicted. Particles with $t_{\rm f} \approx t_{\rm B}$, get
efficiently accreted within a Bondi radius from the core. 

Particles \emph{weakly coupled} to the gas with respect to low-mass cores ($
t_{\rm B} < t_{\rm f}$) are less aided by drag as they get deflected by the
core. As seen in Figure \ref{fig:drift_radius}, a rapid fall-off occurs for
particles with $t_{\rm f} \approx 10^2 t_{\rm B}$. The orbits in the inset show
these particles to be gravitationally scattered, similar to the case were no gas
drag is present.  Here, the physical radius of the core becomes relevant, since
accretion now occurs through gravitational focusing of particles on the core's
surface, which we have not taken into account in Figure \ref{fig:drift_radius}.

\begin{figure*}
  \centering
  \includegraphics{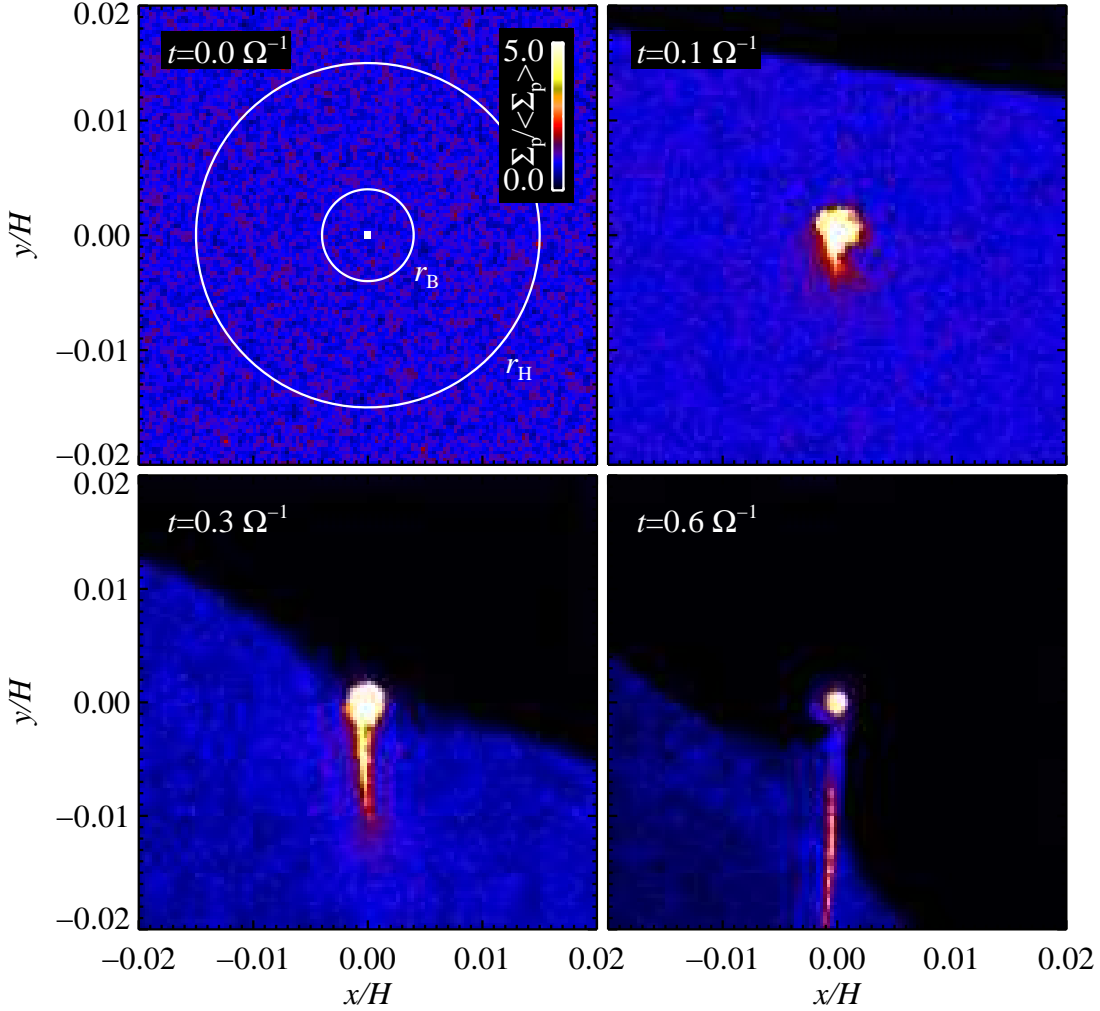}
  \caption{ Accretion of pebbles with $\tau_{\rm f} = 0.1$ by the central core
  ($\mu_{\rm  c} = 10^{-5}$) in the \emph{drift} regime. The color coding shows
  the local particle surface density $\Sigma_{\rm p}$, normalized by the average
  particle density $\left< \Sigma_{\rm p} \right>$, in the simulated shearing
  box with $Z=H_{\rm p}/H=0.01$.  Marked as a white dot, the central seed core
  can be seen in the first panel.  Both the drift and Hill radii are plotted as
  white circles.  The second panel illustrates the drift of the particles
  ($\Delta = 0.05$) and creation of an accreting particle wake. In the third
  panel, particles within the drift radius $r_{\rm d} \approx r_{\rm B}$ are
  accreted.  Particles further out may be carried out of the box by the
  sub-Keplerian gas or by the Keplerian shear. In the final panel the headwind
  has blown most pebbles past the core, with only a minority accreted.}
  \label{fig:overview}
\end{figure*}

\begin{figure}
  \centering
    \includegraphics{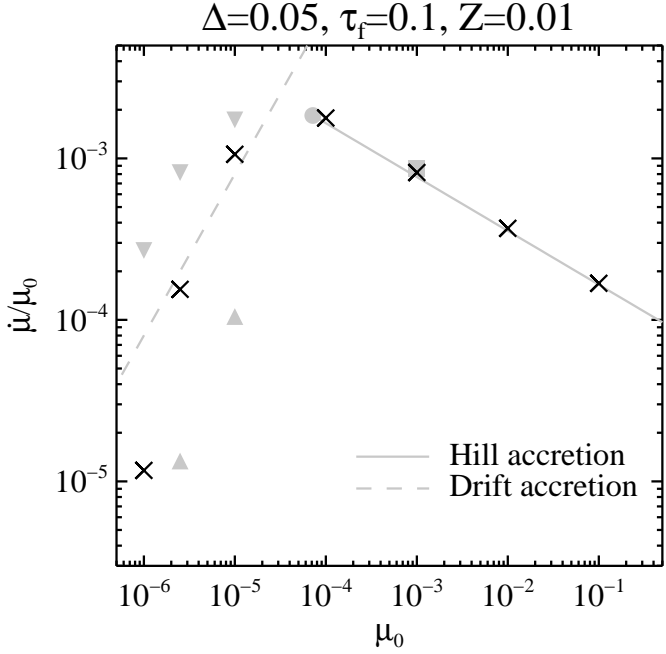}
    \caption{Accretion rate $\dot \mu /\mu_0 $ as function of the initial core
    mass $\mu_0$. Theoretical curves for the branches corresponding to drift and
    Hill accretion are plotted as respectively dashed and full lines in grey.
    The grey full circle marks the transition mass.  Black crosses represent the
    simulated results in a stratified shearing sheet, with $\Delta = 0.05$.
    Triangles correspond to simulations with modified $\Delta$, a grey upwards
    pointing triangle corresponds to $\Delta = 0.07$ and a downwards pointing
    triangle corresponds to $\Delta = 0.03$. In the Hill branch, the position of
    both triangles lie on top of the black crosses and are omitted for clarity.
    The grey square shows the result of simulation {\tt 1e-3\_0.1\_b}, which
    includes the particle backreaction.}
    \label{fig:mu_dot}
\end{figure}

The accretion rate in the \emph{drift accretion} regime is given by 
\begin{equation}
  \dot M_{\rm d} = \pi\rho_{\rm p} r_{\rm d}^2 \Delta v, 
\label{eq:drift_accr_rate}
\end{equation}
when $r_{\rm d}$ is smaller than the particle scale height $H_{\rm p}$.  A
representative simulation in this regime, performed with $\mu_c = 10^{-5}$, is
illustrated in Figure \ref{fig:overview}.  Pebbles drift with a sub-Keplerian
velocity past the core and those entering the Bondi radius, here well inside the
Hill radius, feed the growth of the embryo.  Note that when $r_{\rm d} \approx
r_{\rm b}$, the core growth scales faster than exponential with mass, as $\dot
M_{\rm d} \propto M_{\rm c}^2$.  Figure \ref{fig:mu_dot} shows the accretion
rates calculated from simulations {\tt 1e-6\_0.1, 2.5e-6\_0.1} and {\tt
1e-5\_0.1}.  Particles with friction time $\tau_{\rm f} =0.1$ closely follow the
maximal drift accretion efficiency, 
\begin{equation}
  \frac{\dot \mu_{\rm d}}{\mu_{\rm c}}  = \frac{1}{4} \frac{\rho_{\rm p}}{\rho}
  \frac{\varGamma \mu_{\rm c}}{\Delta^3} \varOmega_{\rm K},
  \label{eq:drift_accr_rate_max}
\end{equation}
with $r_{\rm d} \approx r_{\rm B}$. However the low-mass core in run {\tt 1e-6-0.1}
comes close to the weak coupling limit and sees its accretion rate reduced.

\begin{figure}
  \centering
  \includegraphics{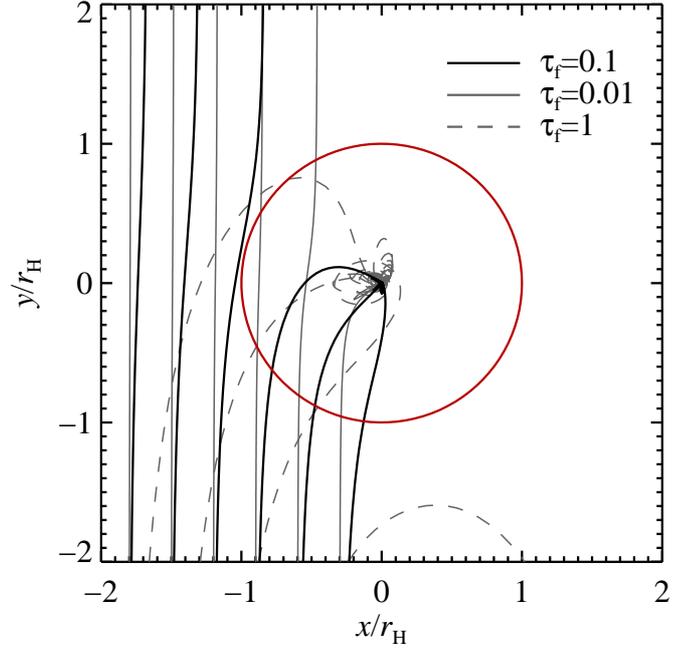}
  \caption{ Trajectories for particles with dimensionless friction time
  $\tau_{\rm f} = 0.01,0.1,1$ obtained from the 2D Hill equations including gas
  drag. Pebbles with $\tau_{\rm f} = 0.1$ and impact parameters below a Hill
  radius efficiently get accreted. Larger particles of $\tau_{\rm f} = 1$ are
  pulled in from wider separations, but cores lose particles on horseshoe
  orbits. Particles strongly coupled to the gas, with $\tau_{\rm f} = 0.01$,
  need close encounters well within the Hill sphere in order to fall onto the
  core. }
  \label{fig:hill_impact}
\end{figure}

\begin{figure}
  \centering
  \includegraphics{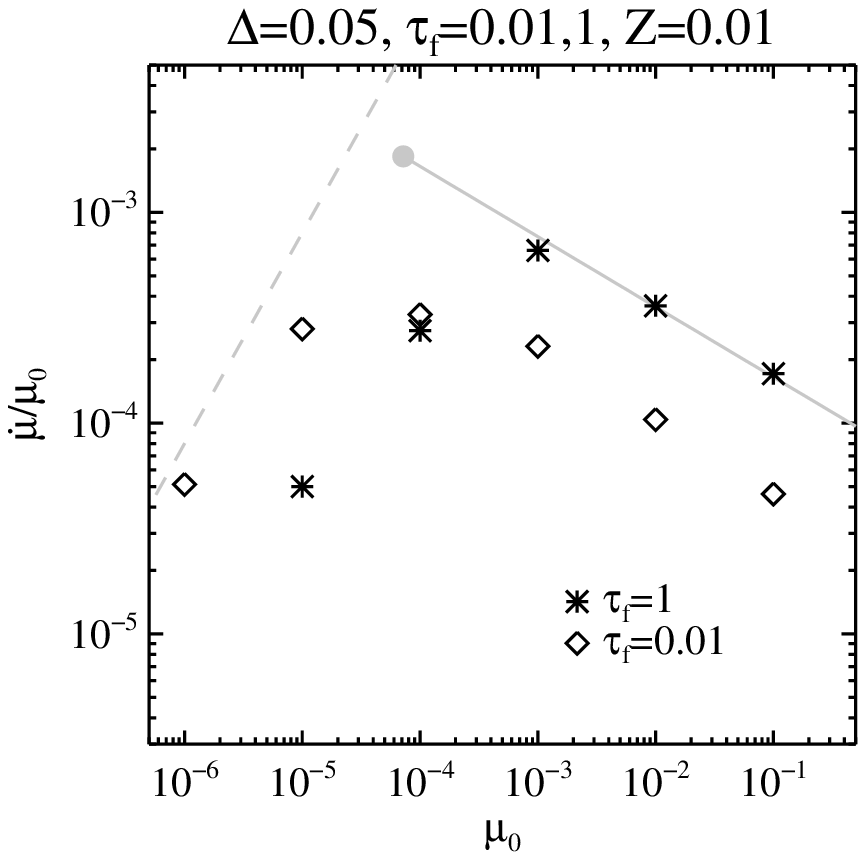}
  \caption{ Normalized accretion rates, $\dot \mu_{\rm c}/ \mu_0$, for different
  particle sizes with friction time $\tau_{\rm f}=1$ and $\tau_{\rm f} = 0.01$,
  as indicated by respectively asterisks and diamonds. The grey curves
  correspond to those shown in Figure\, \ref{fig:mu_dot} and similarly the
  transition mass is indicated by a full grey circle. In the Hill branch, larger
  particles, $\tau_{\rm f}=1$, get accreted as efficiently as  particles with
  friction time $\tau_{\rm f}=0.1$ (see Figure \ref{fig:mu_dot}), but in the
  drift branch they never get accreted at the full drift rate. On the other
  hand, small particles, $\tau_{\rm f}=0.01$, get efficiently accreted in the
  drift branch, but less so in the Hill regime. }
  \label{fig:mu_dot_2}
\end{figure}

We can envisage two effects reducing the accretion rate, if we were to continue
to ignore the stellar tidal field even for higher mass cores.  Firstly, when the
core enters the strong coupling limit, growth slows down to exponential, $\dot
M_{\rm d} \propto t_{\rm B}^{-1} r_{\rm B}^2 \propto M_{\rm c}$.  Secondly, when
the accretion radius becomes comparable to the particle scale height, the
appropriate expression for the accretion rate is given by
\begin{equation}
  \dot M_{\rm d} = 2 r_{\rm d} \Sigma_{\rm p} \Delta v,
\end{equation}
where $\Sigma_{\rm p}$ is the particle column density.  When $r_{\rm d} \approx
r_{\rm B}$, we get exponential growth $\dot M_{\rm d}/ M_{\rm c} =
2\Sigma_{\rm p}/\Delta
v$.

\subsection{Hill accretion}

\begin{figure*}
  \centering
  \includegraphics{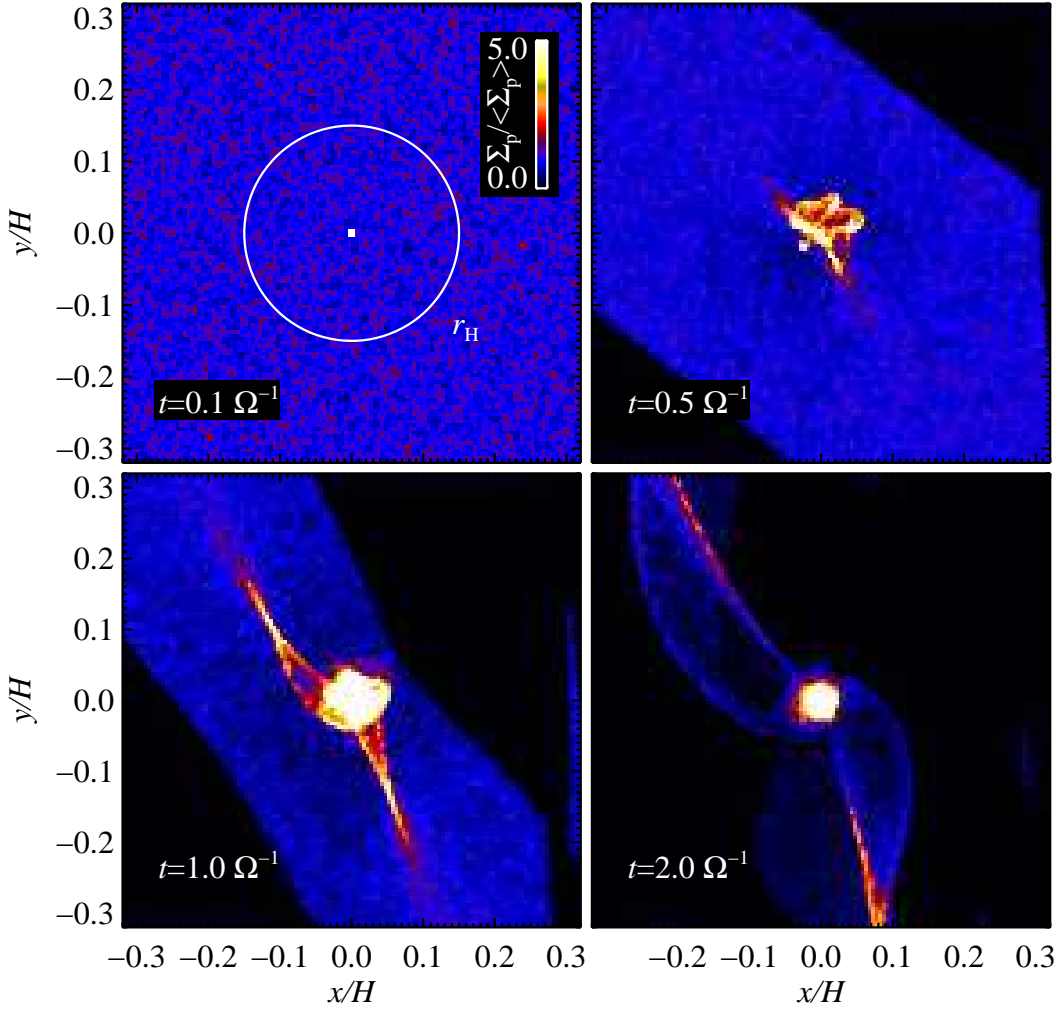}
  \caption{ When the core is massive enough, it can efficiently accrete
  particles with $\tau_{\rm f} = 1$ entering its Hill sphere (indicated by the
  white circle in the first panel).  On this scale, the Keplerian shear
  dominates over the relative velocity difference between the gas and Keplerian
  velocity ($\Delta = 0.05$).   Accretion seems to occur through a particle
  disc, visible after a steady state has set in (panel 2 to 3). In the last
  panel, accretion and Keplerian shear have removed most particles in the box.
  The color coding is similar to Figure \ref{fig:overview}.  The simulation was
  performed in a stratified shearing sheet box with $Z=H_{\rm p}/H = 0.01$.  }
  \label{fig:overview_hill}
\end{figure*}

When the core mass grows to the point where the Bondi radius $r_{\rm B} \propto
M_{\rm c}^2$ is comparable to its Hill radius $r_{\rm H} \propto M_{\rm
c}^{1/3}$ (or identically $v_{\rm H} = \Delta v$ or $t_{\rm B} /t_{\rm f} =
\tau_{\rm f}^{-1}$), it will cross the  transition mass, 
\begin{equation}
  M_{\rm t} = \sqrt{\frac{1}{3}} \frac{\Delta v^3}{G\varOmega_{\rm K}}
  \approx 3 \times 10^{-3} 
  \left( \frac{\Delta}{ 0.05} \right)^3 \left( \frac{r}{{\rm 5\,AU}} \right)^{3/4} {\rm M}_\oplus,
\end{equation}
defined earlier and see a change in pebble accretion mechanism.  The
dimensionless form of the transition mass,
\begin{equation}
  \mu_{\rm t} = 7 \times 10^{-5} \left( \frac{\Delta}{0.05} \right)^3,
\end{equation}
scales as the cube of the headwind parameter $\Delta$ (see Figure
\ref{fig:headwind}).  The Hill radius now sets the maximal impact parameter from
which particles can be accreted.  When $M_c > M_{\rm t}$, pebbles at the edge of
the Hill sphere approach the core with relative velocity 
\begin{equation}
v_{\rm H} \equiv \varOmega_{\rm K} r_{\rm H}.
\end{equation}
The Keplerian shear $v=-(3/2)\varOmega_{\rm K}x$ dominates over the headwind in
the Hill branch, since $ v_{\rm H} /\Delta v = \sqrt{r_{\rm B}/r_{\rm H}}$.  The
inverse Keplerian frequency, $\varOmega_{\rm K}^{-1}$, is the  gravitational
crossing time-scale at the Hill radius, independent of core mass.  For particles
with friction times close to the orbital time-scale ($\tau_{\rm f} =0.1$-$1$),
all particles entering the Hill sphere will be accrete, as illustrated in Figure
\ref{fig:hill_impact}. Here we present particle orbits obtained from  the Hill
equations including drag
\begin{eqnarray}
  \frac{\partial v_x/v_{\rm H}}{\partial t/\varOmega_{\rm K}^{-1}}
  &=& + 2 \frac{v_y}{v_{\rm H}} + 3 \frac{x}{r_{\rm H}}
  - 3 \left( \frac{r}{r_{\rm H}} \right)^{-3} \frac{x}{r_{\rm H}} 
  - \frac{1}{\tau_{\rm f}} \frac{v_x}{v_{\rm H}}\\
  \frac{\partial v_y/v_{\rm H}}{\partial t/\varOmega_{\rm K}^{-1}}
  &=& -2 \frac{v_x}{v_{\rm H}}
   - 3 \left( \frac{r}{r_{\rm H}} \right)^{-3} \frac{y}{r_{\rm H}} 
   - \frac{1}{\tau_{\rm f}} \left( \frac{v_y}{v_{\rm H}} + \frac{3}{2}
   \frac{x}{r_{\rm H}} \right)
  \label{eq:nondimHill}
\end{eqnarray}
where $r = \sqrt{x^2 + y^2}$ is the particle-core distance.  Similar to the
drift case (Section 3.1), when the gravitational deflection time (here
independent of the core mass and $\sim \varOmega^{-1}$) is similar to $t_{\rm
f}$, enough energy will be dissipated during the approach to mediate the
accretion of the pebbles within the Hill sphere.  The accretion rate is then
given by
\begin{equation}
  \dot M_{\rm H} = 2 r_{\rm H} \Sigma_{\rm p} v_{\rm H} \propto M_{\rm c}^{2/3}, 
\end{equation}
since for the core masses under consideration, $M_{\rm c} > M_{\rm t}$, we
accrete the total particle surface density, $\Sigma_{\rm p}$ ($r_{\rm H} >
H_{\rm p}$).  This growth mode is confirmed by our numerical simulations {\tt
1e-4\_0.1, 1e-3\_0.1, 1e-2\_0.1, 1e-1\_0.1} and {\tt 1e-3\_1.0, 1e-2\_1.0,
1e-1\_1.0}. They can be inspected in Figure \ref{fig:mu_dot} and
\ref{fig:mu_dot_2}.  An example of a simulation with a seed mass accreting at
the Hill rate is illustrated in Figure \ref{fig:overview_hill}. Here, particles
of $\tau_{\rm f}=1$ entering the Hill sphere drive the growth of the core of
mass $\mu_{\rm c} = 10^{-2}$. Accretion occurs through a particle disc, as was
previously resolved in high-resolution 2D simulations by
\citet{Johansen_Lacerda_2010}. 

Note that in the classical scenario of planetesimal accretion, one never
captures objects from the full Hill sphere, but only by a fraction $\alpha^{1/2}
r_{\rm H}$, with $\alpha \approx r_{\rm c}/r_{\rm H}$ set by the physical radius of
the core $r_{\rm c}$.  In the terminology of \citet{Rafikov_2011}, slow
accretion of planetesimals between the shear- and dispersion-dominated dynamical
regime, from a part of the particle scale height $H_{\rm p} = v/\varOmega_{\rm K}$
\citep[e.g.][]{Dodson_2009}, goes as approximately $ \dot M \approx \pi \alpha
r_{\rm H}^2 \rho_{\rm p} v \approx \alpha r_{\rm H} \Sigma_{\rm p} v_{\rm H}
\approx \alpha \dot M_{\rm H}$. At 5 AU, this gives a reduction in the accretion
rate of $\alpha \approx 10^{-3} \left( r/5 \rm{\,AU} \right)^{-1}$, for a
standard solid density \citep{Goldreich_2004}.  Accretion of planetesimal
fragments from a thin midplane, as discussed in \citet{Rafikov_2004} is more
efficient, with the accretion rate being proportional to $\sqrt{\alpha} \dot
M_{\rm H}$. At 5 AU, this limits the growth by $\sqrt{\alpha} \approx 3 \times
10^{-2} \left( r/5 \rm{\,AU} \right)^{-1/2}$.  Thus accretion of pebbles at a
rate $\dot M_{\rm H}$  from the entire Hill sphere is extremely efficient,
compared to the classical gas-free case. 

Only the smallest particles we consider, with $\tau_{\rm f} = 0.01$, have an
accretion efficiency that is less than optimal in the Hill branch (Figure
\ref{fig:mu_dot_2}). Similar to the case of the strongly coupled particles in
the drift regime, accretion requires the gravitational deflection time to be
shorter than the friction time, as previously expressed in
Eq.\,(\ref{eq:strongcouple}). The relative velocity for particles approaching
the Hill sphere is set by the Keplerian speed $\Delta v \approx \varOmega r$.
This allows us to rewrite the accretion criterion as 
\begin{equation}
  \varOmega r \frac{r^2}{GM} < t_{\rm f},
\end{equation}
which gives us an effective accretion radius 
\begin{equation}
  r_{\rm eff} \lesssim \tau_{\rm f}^{1/3} r_{\rm H}.   
\end{equation}
In this regime, $\dot M_{\rm H, eff} \propto \tau_{\rm f}^{2/3}$, which compared
to  particles of $\tau_{\rm f} = 0.1$ would give a reduction of the accretion
rate by $\approx 0.2$ , as can be seen from comparing Figure \ref{fig:mu_dot}
and \ref{fig:mu_dot_2}.

\subsection{Influence of headwind reduction and particle feedback}
In the above discussion, we have kept the relative velocity between core and the
gas disc constant at $\Delta = 0.05$. As previously mentioned, we have ignored
the presence of pressure bumps, local extrema in the radial pressure force
resulting in regions of a reduced headwind, as well as extreme orbital distances
where $\Delta$ can change significantly.  While the accretion rate $\dot M_{\rm
H}$ in the Hill regime is insensitive to $\Delta$, the Bondi branch up to the
transition mass $M_{\rm t}$ is not. Figure \ref{fig:mu_dot} illustrates the
effect on changes in $\Delta$ for various core masses ($\Delta = 0.05 \pm
0.02$). For a core mass accreting approximately from the full Bondi branch,
Eq.\,(\ref{eq:drift_accr_rate_max}) indicates that the accretion rate will be
modified by a factor $ (0.05/0.03)^3 \approx 5$ for $\Delta = 0.03$ compared to
accretion with $\Delta = 0.05$. This is in agreement with the measured accretion
rates for $\mu = 2.5 \times 10^{-6}$ ({\tt 2.5e-6\_0.1\_0.03}).  The increase is
reduced for {\tt 1e-5\_0.1\_0.03} where $r_{\rm B}$ grows to $R_{\rm H}$, and
increased for {\tt 1e-6\_0.1\_0.03} where $t_{\rm B}/t_{\rm f}$ grows
sufficiently out of the weak coupling limit. Overall we see that even weak
pressure bumps decreasing $\Delta$ by $0.02$ lead to more rapid accretion.  Vice
versa, for $\Delta = 0.07$ we expect a reduction of the accretion rate by a
factor $ (0.05/0.07)^3 \approx 0.4$ and we measure similar, but lower accretion
rates ({\tt 1e-5\_0.1\_0.03, 2.5e-6\_0.1\_0.03}).

\begin{figure*}
  \centering
  \includegraphics{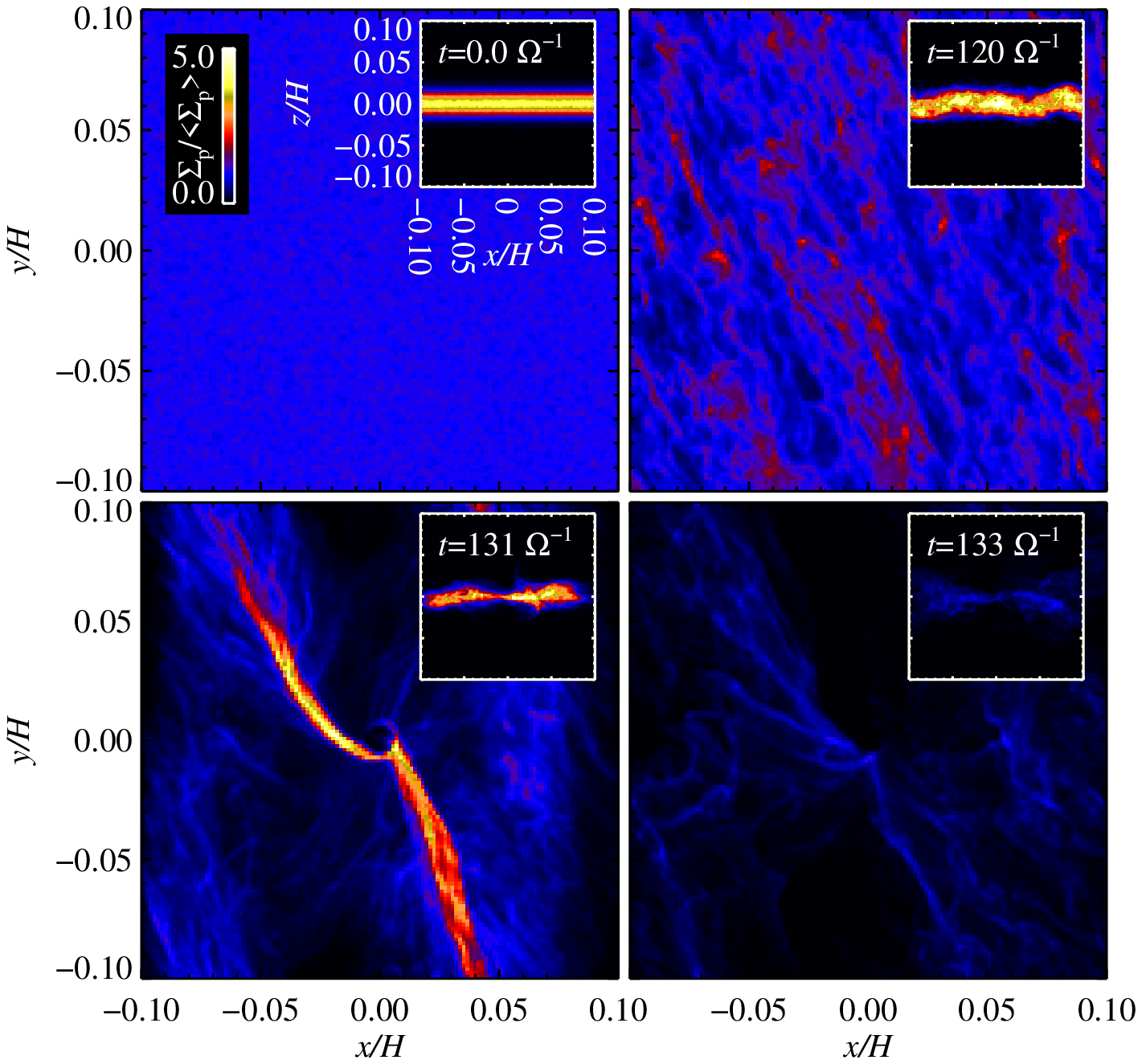}
  \caption{ Including the backreaction friction force on the gas, a core of $\mu
  = 10^{-3}$ accretes particles of $\tau_{\rm f} = 0.1$ (third and last panel),
  even in presence of turbulence caused by the streaming instability.  The
  turbulence has first been given $20$ orbits ($\sim$$126\,\varOmega^{-1}$) to
  saturate (first and second panel), before the core is inserted.  From the full
  Hill sphere, $r_{\rm H}/H \approx 0.07$, pebbles are attracted in a prograde
  motion to the core (third panel).  As before, the color bar in the first panel
  gives the particle surface density. The insets show the azimuthally averaged
  particle density and vertical extent of the particle layer. For clarity, in
  the insets the color coding covers a twice as wide range in particle
  overdensity, compared to the surface density plots.  }
  \label{fig:overview_hilli_back}
\end{figure*}

In simulation {\tt 1e-3\_0.1\_b}, we have departed from a smooth gas velocity
profile, by including friction on the gas and letting turbulence develop by the
streaming instability.  After approximately $20$ orbits, we place the seed core
mass in the center of the simulated domain (Figure
\ref{fig:overview_hilli_back}). The measured accretion rate does not deviate
measurably from the case not including particle backreaction, as can be seen in
Figure \ref{fig:mu_dot}. This indicates that our results are robust for the Hill
branch, even in a turbulent environment. More simulations should be carried out
in the future to verify the validity in the drift regime. However, this requires
very high resolution in order to resolve both the Bondi radius and the streaming
instability wavelength simultaneously.

\section{Implications for gas giant growth}

Having numerically confirmed the pebble accretion rates for low to high core
masses, we can extrapolate our results and find the time necessary to grow a core
to the critical mass needed to attract its gaseous envelope. 

In the drift regime, the accretion rate implies a growth time-scale of 
\begin{equation}
  \Delta t_{\rm d} = \int_{M_0}^{M_{\rm c}} \dot M_{\rm d}^{-1}dm
  \approx \frac{\Delta v^3}{\pi\rho_{\rm p} G^2} M_0^{-1}
\end{equation}
to reach a core mass $M_{\rm c}$ from an initial seed mass $M_0 \ll M_{\rm t}$.
This lower limit, since we assume optimal accretion from the full Bondi radius,
sets the time until the accretion rate blows up hyperbolically as $M_{\rm c}
\propto \left( \Delta t_{\rm d} -t \right)^{-1}$ and thus does not depend on the
final mass we wish to reach. However, from our numerical results we know that
this growth is not sustained and turns off to the Hill branch. The
characteristic time to reach the end point for drift accretion, the  transition
mass $M_{\rm t}$, is given by 
\begin{equation}
  \Delta t_{\rm d} 
  \approx 8 \times 10^6 \left( \frac{\Delta}{0.05} \right)^3 
  \left(\frac{\rho_{\rm p}/\rho_0}{0.01}  \right)^{-1}
  \left( \frac{M_0 }{ 10^{-5} \rm{ M}_\oplus} \right)^{-1} 
  \left( \frac{r}{5 \rm{\, AU}} \right)^2
  \rm{yr },
\end{equation}
which at $5$ AU is comparable to the gas disc lifetime. If particles sediment to
the midplane, the ratio of the particle to the gas density, $\rho_{\rm
p}/\rho_0$, would be of order unity. However, past $0.5$ AU, a small seed core
mass, $M_0 = 10^{-5}$ M$_\oplus$,  accretes non-sedimented particles of size
$\tau_{\rm f} < 0.01$ most efficiently.  For reasonable values of the local
headwind, $\Delta = 0.03$ - $0.07$, growth is too slow to form cores large
enough to enter the Hill accretion regime, and without pressure bumps $\Delta$
is even larger at wide stellar separations.  

Hill accretion on the other hand, has a growth time-scale of 
\begin{equation}
  \Delta t_{\rm H} = \int_{M_{\rm t}}^{M_{\rm crit}} \dot M_{\rm H}^{-1}dm 
  \approx \frac{3^{5/3}\varOmega_{\rm K}^{1/3}}{2G^{2/3}\Sigma_{\rm
  p}}M_{\rm{crit}}^{1/3},
\end{equation}
which is only weakly dependent on the critical mass for gas envelope attraction
$M_{\rm crit}$ and independent of the transition mass, when $M_{\rm t} \ll
M_{\rm crit}$.  The core growth when accreting pebbles is fast in this regime,
at $5$ AU the critical mass is reached after
\begin{equation}
  \Delta t_{\rm H} \approx 4\times10^4 \left( \frac{M_{\rm{crit}}}{10 {\rm\,
  M}_\oplus} \right)^{1/3} \left( \frac{r}{ 5{\rm\, AU}} \right) \rm{\, yr}.
\end{equation}
Furthermore, the growth time-scale $\Delta t_{\rm H}$ scales linearly with
orbital distance $r$, as opposed to quadratic in the drift regime. This makes
core formation possible in distant regions of the protoplanetary disc.  Also,
note that Hill accretion rate is maintained for a single particle size with
friction times $\tau_{\rm f} \sim 0.1$-$1$, independent of the core mass, as
opposed to the Bondi regime where one unrealistically needs to maintain $t_{\rm
B} \approx t_{\rm f}$ to maintain the maximal accretion rate. 

\begin{figure}
  \centering
  \includegraphics{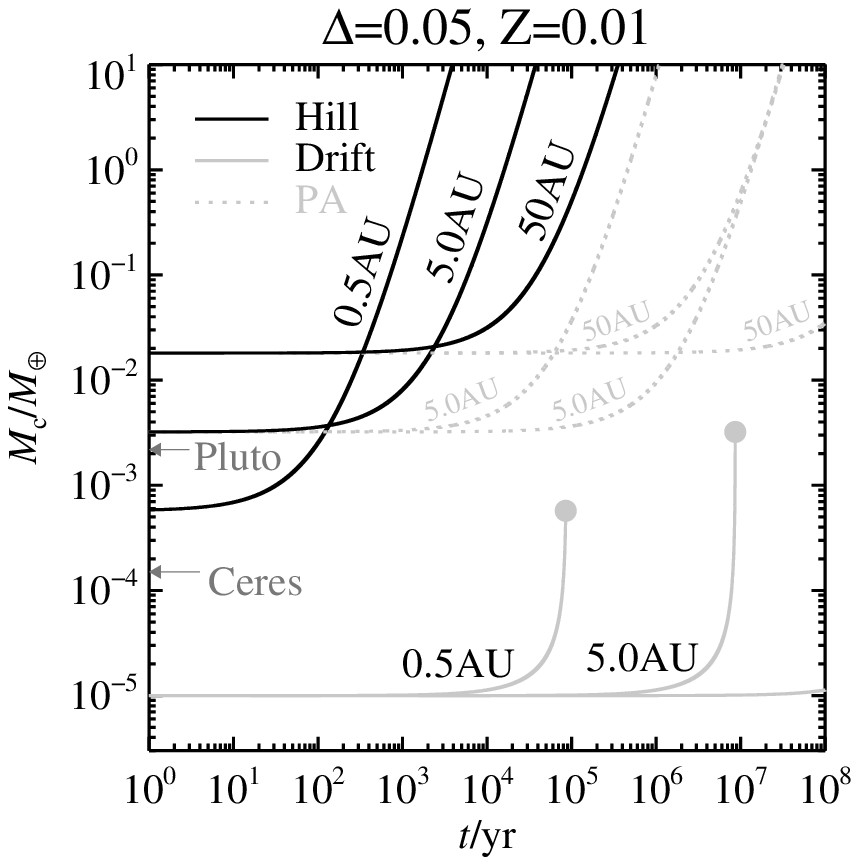}
  \caption{Core growth as function of time, plotted for various orbital
  distances ($0.5,5$ and $50$\,AU). The drift branch, marked by grey solid
  lines, assumes an initial core mass of $M_0 = 10^{-5} M_\oplus$ and $\Delta =
  0.05$. The drift growth continues until the transition mass $M_{\rm t}$ is
  reached (marked by a full grey dot). Accretion continues through the more
  efficient Hill branch, drawn in black. For clarity, we start the Hill growth
  from the transition mass at time $t=0$\,yr, instead of continuing from the
  time where drift accretion comes to a halt. The masses of Ceres and Pluto
  (located at respectively $2.7$ and $39$ AU) are marked on the vertical axis
  for reference.  The grey dotted curves correspond to classical planetesimal
  accretion (PA), where the faster growth corresponds to 2D accretion of
  planetesimal fragments \citep{Rafikov_2004} and the slower to 3D accretion of
  planetesimals \citep[e.g.][]{Dodson_2009}.  Note that drift accretion
  time-scale at $50$\,AU takes more than $10^8$\,yr and its transition mass
  point is not plotted.}
  \label{fig:M_t}
\end{figure}

Figure \ref{fig:M_t} shows the core growth in both regimes and the dependency on
the orbital distance. We conclude that fast core growth is possible through
pebble accretion, provided that the initial seed mass for the core is above the
local transition mass.  A sufficiently large embryo can only be grown by drift
accretion in pressure bumps with low headwind, $\Delta \lesssim 0.05$, or be the
result of planetesimal formation by gravitational collapse after concentration
by e.g.~streaming instabilities (see also Figure \ref{fig:muvsr}).  It is
interesting to note that both Ceres and Pluto have less than the critical mass
needed for fast Hill accretion, which might explain why they failed to grow to
gas or ice giants. Indeed, one can make the assumption that only those
planetesimals that formed early enough in the high-mass tail of the initial
planetesimal mass distribution could serve as the seed for gas-giant cores.

\begin{figure}
  \centering
  \includegraphics{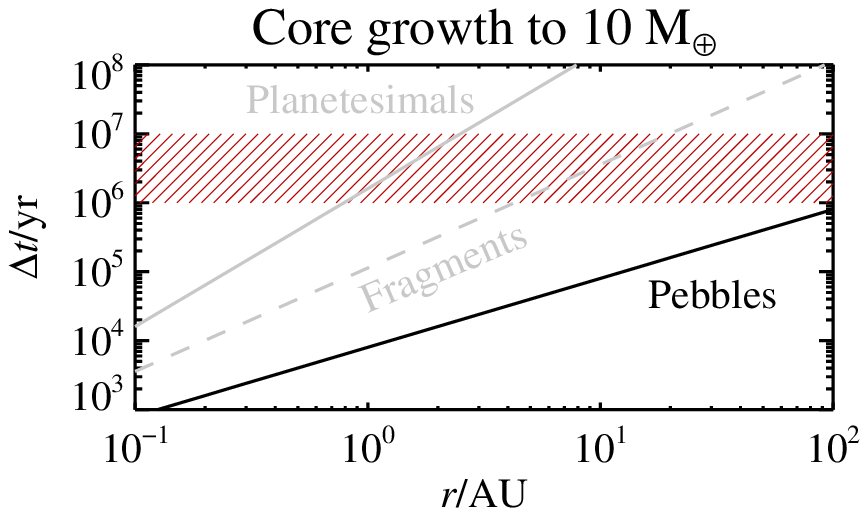}
  \caption{ Time needed for core growth up to 10 M$_\oplus$ at various locations
  in the disc. The solid black line gives the formation time of the core for
  pebble accretion in the Hill regime, while grey lines give the time needed to
  form the critical 10-Earth-mass core by planetesimal accretion. The dashed
  grey line represents planetesimal fragment accretion from a thin midplane
  layer, as studied by \citet{Rafikov_2004}.  The red shaded area shows the
  approximate time interval in which the protoplanetary disc loses its gaseous
  component and encompasses for example the estimated age of gas giant LkCa 15b
  \citep{Kraus_2012}. Core formation needs to occur before this time.}
  \label{fig:growth_time} 
\end{figure}

As an illustration of the rapid core growth by pebble accretion in the Hill
regime, we compare it to the core growth time for planetesimal accretion in
Figure \ref{fig:growth_time}. As discussed in Section 3.2, the inability to
accrete solids from the entire Hill sphere, as opposed to pebble accretion,
leads to significant longer core formation times, in conflict with the observed
dissipation time of protoplanetary discs.

\section{Discussion}
We discuss here the assumptions and limitations of our results.

{\it Midplane layer thickness.} One component of the pebble accretion scenario
is the presence of a thin particle disc ($H_{\rm p} = 0.01 H$). This low
particle scale height is expected from turbulence driven by streaming
instabilities, independent of particle size, as discussed in Section 2.  A
moderately higher particle scale height, as may be the case for turbulence
caused by the magnetorotational instability, can result in a situation where
$r_{\rm H} < H_{\rm p}$ past the transition core mass.  This would result in a
temporarily reduced accretion rate, by a factor $r_{\rm H}/H_{\rm p} = (H_{\rm
p}/H)^{-1}(r_{\rm H}/H) = (1/3)^{1/3}(H_{\rm p}/H)^{-1}\mu^{1/3} $, until the
Hill radius grows beyond the particle scale height.

{\it Particle size.} The assumption of a single particle size in our simulations
can be criticised, but as discussed in the introduction, observations of
protoplanetary discs allow a large fraction of the solid mass to reside in the
particle size range that we consider, $\tau_{\rm f} = 0.01$-$1$
\citep{Wilner_2005}.  A large abundance of particles larger than pebbles is not
expected from coagulation models \citep{Blum_2008, Brauer_2008,Windmark_2012}.
However, as particles approach the core their icy component might sublimate; as
friction would heat the particles, especially when a denser envelope starts
forming around the core. It would be interesting to take this size-diminishing
effect into account in a further investigation. On the other hand, particles
might grow larger. In higher metallicity environments streaming instabilities
become so effective in clumping solid material that one can fear particles to
grow past the pebble size. However, we do not see this particle clumping in our
simulations including the gas drag backreaction at the metallicity we consider
($Z=0.01$). Strong clumping requires $Z \gtrsim 0.02$ \citep{Johansen_2009_met,
Bai_2010}.

{\it Gas structure.} For the lower seed masses discussed in the paper, we
previously argued (Section 2) that the gas density changes around the core are
small.  In the Hill regime, the ratio $v_{\rm esc}^2/ c_{\rm s}^2 \approx 2.3
\times 10^2 \mu^{2/3} (r/{\rm AU})$ (in the MMSN for standard solid density) can
exceed unity for the highest core masses and the effects of an envelope should
be taken into account.  But, as also argued by \citet{Ormel_2010}, even if the
direction of the flow moderately changes on scales within the Bondi radius due
to stratification near the core, only particles with $ t_{\rm f} \ll t_{\rm B}$
could be affected by it. Since these particles are too strongly coupled to the
gas for accretion to take place in the first place (strong coupling limit),
ignoring the core's feedback on the gas is justified.

{\it Keplerian orbits.} In our analysis we assumed the core to be on a circular
Keplerian orbit.  The relative velocity between the core and the gas in
Keplerian rotation could be significantly modified if competing cores would get
excited by repeated close passages. However, as opposed to classical
planetesimal growth, in our scenario gas damps the small particles and dynamical
friction prevents the excitation of larger bodies, similar to the oligarchic
growth regime.  We do ignore gas-driven type-I migration of the core, important
for core masses over $0.1$ M$_\oplus$ \citep{Tanaka_2002}.  

{\it Random particle speed.} In our simulations, particles approach the core in
equilibrium with the gas flow. Particle interactions with the core last at most
of the order $\varOmega_{\rm K}^{-1}$, as in the Hill regime. Small particles
($\tau_{\rm f} \lesssim 1$) are coupled to the gas on similar time-scales. The
passage of the core is quickly erased for the non-accreted particles, even when
ignoring radial drift and turbulent diffusion. The core only catches up with the
deflected particles after approximately  $t_{\rm pass} = 2\pi r \Delta v^{-1} =
(2\pi/\varOmega_{\rm K}) (r/H) \Delta^{-1} \approx 10^4 \left( r/{\rm AU}
\right)^{-1/2} \varOmega_{\rm K}^{-1}$.  However, in the Hill accretion regime
all particles that can be deflected are accreted, and it is the radial drift and
diffusion of particles that fill up the feeding zone. Diffusion can be rapid,
since the diffusion time associated with closing the Hill sphere $t_{\rm H}
\varOmega \sim R_{\rm H}^2/(\delta_{\rm t} H^2 ) \sim (1/3)^{2/3} \delta_{\rm
t}^{-1} \mu^{2/3}$, is of order unity for a protoplanetary disc with
$\delta_{\rm t} =0.01$.

{\it Particle drift.} When the drag force responsible for radial drift is too
small, particles could get trapped in mean motion resonances with the core.
\citet{Weidenschilling_1985} studied large, $\tau_{\rm f} \geq 1$, particles in
the Stokes drag regime, and argued that particles smaller than these sizes feel
large enough drag forces to escape resonant trapping around a Jupiter-mass
planet at $5$ AU. As shown by \citet{Tanaka_1997} inclusion of mutual
planetesimal interactions breaks down the resonances, but dust gap formation
still occurs for large planetesimals, where gas drag changes the semi-major axis
of the the planetesimals after scattering with the protoplanet. The maximal
particle size unaffected by particle trapping seems approximately inversely
proportional to the planet's mass, which is also seen in simulations performed
by \citet{Paardekooper_2007}.  In fact, \citet{Weidenschilling_1985} argue that
small pebbles are the only size that can be accreted by the core, since trapped
larger planetesimals get dynamically excited and will be ground to fragments,
which in their turn are capable of escaping the resonance. This picture is
confirmed in simulations performed by \citet{Levison_2010}. 

Dust gaps can open up before the core is massive enough to create a gap in the
gas disc itself \citep{Paardekooper_2006}.  \citet{Muto_2009} analytically show
that the core has to be over a critical mass, 
\begin{equation}
  \mu_{\rm c} > \Delta \left( \frac{H}{r} \right)^{-1} \approx 1,
  \label{eq:gap_crit}
\end{equation}
for particles of $\tau_{\rm f} \leq 1$ in order for a dust gap to emerge. Past
$r \approx 1$ AU, $\mu \approx 1$ is consistently above $10$ M$_\oplus$, the
critical core mass for gas and ice giants (see Figure \ref{fig:muvsr}).
Particles thus always drift radially fast enough to replenish the feeding zone
of the core. Indeed, if the drift rate is set by $R_{\rm d} = 2\pi r \Delta v
\Sigma_{\rm p}$, the requirement $R_{\rm d} \geq \dot M_{\rm H}$ recovers the
above criterion, Eq.\,(\ref{eq:gap_crit}).  At the same time, as pointed out by
\citet{Ormel_2011}, the particle drift can also be responsible of clearing up
the entire reservoir of available pebbles in the disc.

{\it Terrestrial planet formation.} Growth at small orbital distances, $r <
5$\,AU, is remarkably rapid in the pebble accretion model. Formation of rocky
planets and possibly in situ formation of gas-giant planets in the terrestrial
planet region seems problem-free from the perspective of the accretion rate. The
growth time-scales for both the drift and Hill accretion branch shrink to
approximately $10^5$ yr at Earth-like separations from the host star. This could
indicate that even terrestrial planet formation occurs rapidly during the
gaseous disc phase. However, closer to the star the amount of material in an
annulus of Hill-radius-width  is small and the isolation mass by gap formation
is lower.  Also the optimally accreted particle size is large, around $10$ cm,
and ices are not available.

\section{Conclusions}
In this paper we have demonstrated that accretion of pebbles makes rapid
formation of gas-giant cores possible.  The growth time-scale to reach the
critical core mass for gas accretion is reduced by  three orders of magnitude at
$5$ AU and four orders of magnitude at $50$ AU, compared to the planetesimal
accretion rate in between the shear- and dispersion-dominated dynamical regime.
Compared to accretion of planetesimal fragments from a thin layer the formation
time is shortened by approximately a factor 30  at $5$ AU and a factor 100 at
$50$ AU.  This is further support for the core accretion scenario, because cores
can form by pebble accretion before gas dissipation after $1$-$10$ million
years, even at large orbital radii.  

We can summarise the main numerical results as follows.  Our simulations show
gas drag to be a necessary ingredient for fast pebble accretion by the growing
core.  Omnipresent pebbles, particles with friction time around $\tau_{\rm f}
\approx 0.1$, are ideally suited for core growth. They are weakly enough bound
to the gas to feel the gravitational pull from the core, but strongly enough to
deposit their kinetic energy through drag forces, when passing the core.
Low-mass cores, cores below the transition mass corresponding to a body of
radius larger than approximately $1,000$ km, can accrete small particles
drifting with the sub-Keplerian gas velocity past the core, but this process is
slow, even in pressure bumps with reduced headwind and particle settling in a
thin mid-plane layer. However growth in this regime could be important for a
seed planetesimal formed just below the transition mass, where the accretion
rate is high.  Higher-mass cores can efficiently attract pebbles from the full
Hill sphere, as was found by \citet{Johansen_Lacerda_2010} and
\citet{Ormel_2010}. In this regime, the optimally accreted particle size is
independent of the core mass, in contrast to the drift regime, where the
particle size with the highest accretion rate increases linearly with radius of
the growing core.

For the pebble accretion mechanism to be rapid,  a significant fraction of the
solid density needs to be in the form of pebbles close to the midplane, and some
planetesimals need to form with sizes of $1000$ km or larger. Theoretical models
of planet formation show that these large seeds of approximately Ceres-size can
form by self-gravity after clumping by the streaming instability
\citep{Johansen_2007_nature, Johansen_2012}, so the pebble accretion scenario
fits well with the formation of planetesimals by self-gravity.

The conditions for fast core growth are supported by observations.  A large
reservoir of pebbles is inferred in observations of many young protoplanetary
discs \citep{Wilner_2005, Rodmann_2006}.  Additionally, studies of the
collisional evolution of the asteroid belt show that large asteroids must have
formed early when gas was still present \citep{Morbi_2009}.  The early
disappearance of mm-dust in protoplanetary discs on time-scales shorter than 1
Myr \citep{Lee_2011} can be contributed to fast particle growth and rapid core
formation. In fact, we see that gas giants form both rapidly \citep{Kraus_2012}
and at large orbital radii \citep{Marois_2010}.  Our results predict that ice
and gas giant planets, detectable with direct imaging surveys, will be abundant
around young ($\sim$$1$ Myr) stars.

Further exploration of the pebble accretion mechanism by simulations with
particle backreaction, around higher or lower metallicity discs in  larger
simulation domains, are needed to show the robustness of rapid core growth.
Preferably, global simulations should be developed, including gap formation,
radial drift, mean motion resonances and multiple cores, in order to get a full
overview on the implications on fast core growth.  When the core mass starts to
approach the critical 10 Earth-masses for envelope attraction, we see a prograde
particle disc emerge. Studying these discs, possibly the birthplaces of the
regular satellites of gas giants, in conjunction with the emergence of a dense
envelope around the core, will teach us more about the early growth and gas
accretion thermodynamics of gas giants. 

Pebble accretion provides us with a viable pathway to rapid
formation of gas-giant and ice-giant cores. The necessary first step of the core accretion scenario can 
occur even at wide stellar separations, well within the lifetime
of gaseous protoplanetary discs.

\begin{acknowledgements} Computer simulations were performed at the Platon
  system of the Lunarc Centre for Scientific Computing at Lund University.  Part
  of this work was supported by the Royal Physiographic Society.  We thank
  Alessandro Morbidelli, Matthew Kenworthy and Chris Ormel for insightful
  discussions. The authors thank an anonymous referee for helpful comments.
\end{acknowledgements}

\end{document}